\documentclass{article}

\usepackage[utf8]{inputenc} 
\usepackage[T1]{fontenc}    
\usepackage[hidelinks]{hyperref}       
\usepackage{url}            
\usepackage{booktabs}       
\usepackage{amsfonts}       
\usepackage{nicefrac}       
\usepackage{microtype}      
\usepackage{xcolor}         

\usepackage{amsmath}
\usepackage{amsthm}
\usepackage{amsfonts}
\usepackage{graphicx,psfrag,epsf}
\usepackage{enumerate}
\usepackage{natbib}
\usepackage{hyperref}
\usepackage{url} 
\usepackage{bbm}

\newcommand{\Var}[2]{\, {\rm Var}_{ #1 }\left[ #2 \right] }
\newcommand{\EE}[2]{\mathbb{E}_{ #1 } \left[ #2 \right]}
\newcommand{\Ee}[1]{\mathbb{E}\left[ #1 \right]}
\newcommand{\Sum}[2]{\sum\limits_{#1}^{#2}}

\newtheorem{assumption}{Assumption}
\newtheorem{corollary}{Corollary}
\newtheorem{proposition}{Proposition}
\newtheorem{definition}{Definition}
\usepackage{xr}
\newcommand{\add}[1]{\textcolor{black}{#1}} 

\newcommand{\blind}{0}

\def\spacingset#1{\renewcommand{\baselinestretch}%
{#1}\small\normalsize} \spacingset{1}

\begin{document}

\if0\blind
{
\title{\bf Online detection of forecast model inadequacies using forecast errors}
  \author{Thomas Grundy\hspace{.2cm}\\
    STOR-i Centre for Doctoral Training, Lancaster University, UK\\
    Peak.ai, Manchester, UK \\
    and \\
    Rebecca Killick \\
    School of Mathematical Sciences, Lancaster University, UK \\  \texttt{r.killick@lancaster.ac.uk}\\
    and \\
    Ivan Svetunkov \\
    Department of Management Science, Lancaster University, UK}
  \maketitle
} \fi

\if1\blind
{
  \bigskip
  \bigskip
  \bigskip
  \begin{center}
    {\LARGE\bf Online detection of forecast model inadequacies using forecast errors}
\end{center}
  \medskip
} \fi

\begin{abstract}
In many organisations, accurate forecasts are essential for making informed decisions for a variety of applications from inventory management to staffing optimization. Whatever forecasting model is used, changes in the underlying process can lead to inaccurate forecasts, which will be damaging to decision-making. At the same time, models are becoming increasingly complex and identifying change through direct modelling is problematic. We present a novel framework for online monitoring of forecasts to ensure they remain accurate. By utilizing sequential changepoint techniques on the forecast errors, our framework allows for the real-time identification of potential changes in the process caused by various external factors. We show theoretically that some common changes in the underlying process will manifest in the forecast errors and can be identified faster by identifying shifts in the forecast errors than within the original modelling framework. Moreover, we demonstrate the effectiveness of this framework on numerous forecasting approaches through simulations and show its effectiveness over alternative approaches. Finally, we present two concrete examples, one from Royal Mail parcel delivery volumes and one from NHS A\&E admissions relating to gallstones.
\end{abstract}

\noindent%
{\it Keywords:}  structural break, changepoint, sequential change.
\vfill

\section{Introduction}\label{sec:Introduction}
Accurate forecasting is crucial for planning and marketing decisions for many companies. For example, within inventory management, the optimization of stock levels crucially depends upon accurate supply and demand forecasts \citep[see, for example reviews by][]{Fildes1985, Fildes2008, Syntetos2016}.
If forecasts (or prediction intervals based upon them) start performing poorly this can have severe consequences e.g., product shortages in the retail sector. Hence, it is crucial to quickly identify when forecasting models start deteriorating in performance, in order to react and make necessary adjustments.

A common cause of forecasts becoming inaccurate is due to a changepoint (also known as a structural break) occurring in the underlying process.
Throughout we call the data generated from the underlying process the raw data. After a changepoint in the raw data, the current forecasting model may become inappropriate and the model needs rebuilding. Without the knowledge of changepoints, it is disputed how long forecasting models can be used before updating \citep{Krasheninnikov2018}.
The solution of frequent model updating ignoring potential changepoints is poor data utilisation and may use a mix of pre- and post-change data, therefore damaging the resulting forecasts \citep{Chapman2020}.

A common approach to identify changes is to add a changepoint component to the modelling paradigm; this is easier in some paradigms than others. Even when it is clear how to add changes to our modelling paradigm, if our model is sophisticated (for example containing seasonality, temporal dependence, exogenous variables and/or trends) then our ability to identify a change quickly deteriorates with increasing complexity. In this paper, we propose a framework for sequential changepoint detection within forecast errors to detect if a model becomes inaccurate. As mentioned, the most likely cause for forecasts becoming inaccurate is a change in the raw data being forecast. A change in the mean of the raw data could result in bias in the forecast and this would show as a mean change in the forecast errors. A change in the variance of the raw data, may not cause bias in the forecasts but could lead to miss-calibrated prediction intervals. This change would be reflected by a variance change in the forecast errors. Hence, by detecting changes in the forecast errors, we can automatically identify when the forecasting model needs to be reconstructed and/or re-estimated. This eliminates the need for manual checks of the forecasting models and allows practitioners to focus on other tasks.

Sequential changepoint analysis aims to detect changes in a data generating process in an online manner. As more data becomes available, they are sequentially checked to determine if the distribution of the data has changed. Sequential changepoint analysis dates back to \cite{Page1954} and traditionally aims to identify changes in the mean of the data generating process under the assumption of independent observations and constant variance, see \cite{Basseville1993}, \cite{Csorgo1997} and \cite{Tartakovsky2014} for overviews. Our approach to sequential changepoint detection stems from the work of \cite{Chu1996}, where sequential changepoint procedures are considered in two phases. In phase 1, we assume a fixed amount of data is readily available and is generated from the same distribution; we call this the training data. The training data is used to estimate the data generating process.
Sequential monitoring begins in phase 2; i.e, when the forecasting model goes live. As more data becomes available we make a decision whether the new data is still being generated from the same distribution as Phase 1. If the new data significantly differs from the data in Phase 1, then we deem a changepoint to have occurred and flag that model recalibration is required. Hence, here our model is trained once using the fixed amount of training data and is not re-trained; either a changepoint is detected and the monitoring stops or the monitoring continues indefinitely.

The work of \cite{Chu1996} has been extended in various ways including sequential monitoring of linear models \citep{Horvath2004}; autoregressive processes \citep{Gombay2009}; and financial time series \citep{Aue2009}. Here we utilize sequential changepoint methods to detect changes in forecast errors to quickly identify when forecasts become inaccurate. In particular, we examine one-step-ahead forecast errors, which, within models with an additive error structure, are the model residuals. There has been previous work utilizing the residuals of a model to detect changepoints. In particular, \cite{Horvath2004} use the residuals from estimated linear models to detect changes in regression co-efficients. Furthermore, \cite{Aue2015} use the residuals from estimated ARMA processes to detect changes in mean, variance or regressive parameters. Moreover, within the streaming classification literature there are numerous methods that use classification errors to detect when changes have occurred \citep{Gama2004, Ross2012}. In the offline changepoint setting where all data is available a priori, \cite{Robbins2011} proposed using residuals within the CUSUM procedure and showed that this gave an improved performance over using the raw data when temporal dependence was present. All the methods mentioned above, assume a specified, known forecasting model. Our framework is far more general as any forecasting approach could be used, including model-based, machine learning and expert judgement forecasts - we only require the forecast errors and some assumptions on their properties.

The novelty in our work is two-fold. From a forecasting viewpoint, by combining forecasts and sequential changepoint detection we create a novel framework for automatically identifying when a forecasting model needs re-evaluating. Moreover, we show theoretically how some common changes in raw data (which are a likely cause of inaccurate forecasts) manifest in the forecast errors and therefore can be detected within our framework. From a sequential changepoint viewpoint, this approach can provide an improved performance over model-based sequential changepoint techniques. If the data generating process is sophisticated, then large amounts of data are needed to get an adequate model. Hence, after a change has occurred, we would expect a longer detection delay before signalling a change as more data is required to model the post change distribution. In contrast, performing sequential changepoint detection on the forecast errors is a much simpler task as the post-change modelling issues are eradicated by the forecasting model.

The paper proceeds as follows.
In Section \ref{sec:Methodology} we introduce our framework for monitoring forecast errors and show theoretically how different changes in potentially sophisiticated models manifest in these forecast errors. Readers unfamiliar with sequential changepoint detection can reference Section A in the Supplementary Material for background details.  Section \ref{sec:Examples}, shows how two common forecasting models fit into our framework before Section \ref{sec:Simulation}, demonstrates the improved performance of monitoring the forecast errors over monitoring the raw data in a number of simulated examples. Finally, Section \ref{sec:Application} shows the effectiveness of our method in two applications, one to NHS A\&E admissions data, and the other to Royal Mail delivery volumes,
before Section \ref{sec:Conclusion} gives concluding remarks and suggests some future research directions.

\section{Monitoring Forecast Errors}\label{sec:Methodology}
The main aim of this paper is to introduce a framework that can quickly identify if a forecasting model becomes inaccurate. A common cause of inaccurate forecasts is a change in the potentially sophisticated raw data. For simple data, we can detect mean changes by performing sequential changepoint detection on the data, namely Page's CUSUM detector, as described in Section A of the supplementary material. We can also adapt this detector to detect variance changes by monitoring the (centered) squared data as described in \cite{Inclan1994}. Yet, it is increasingly common for data to exhibit more sophisticated patterns than this setting allows. Some common examples of such structure include seasonality, exogenous relationships, and trend with the data. Moreover, if the data exhibits temporal dependence then convergence to the limiting distributions in Section A are slow and the performance of the detector deteriorates. There has been some work to identify changepoints in some specific models such as data exhibiting trend \citep{Horvath2004} and ARMA models \citep{Aue2015}, however, our work is more general and encompasses these models and more by allowing for a generic forecasting model\add{, which could include artificial neural networks, a blend of experts, or ensemble approaches}.

Alternatively, model-based sequential changepoint techniques could be used for more sophisticated models. These fit a model to the data before and after a potential changepoint and use a likelihood ratio style approach to determine if a changepoint has occurred. Even if the model is known this approach has a number of drawbacks. Firstly, we need to have at least as many observations as parameters in the model. This means if we have seasonality in the model, we require at least one full season worth of data just to fit the post-change model; this can greatly increase the detection delay before a change is identified, especially in \add{high-frequency processes with low-frequency components e.g., daily data with yearly seasonality}. Additionally, the computation time required to fit more sophisticated models increases with complexity.  Furthermore, as we get more data points, there are more potential changepoint locations to check for a change, thus further increasing the computational time. Finally, the choice of threshold in sophisticated models is non-trivial, so identifying changepoints while controlling the false alarm rate is challenging.

Our framework bypasses the above issues by monitoring forecast errors, instead of the raw data, for changepoints. The intuition behind this is that the forecast model should capture the sophistication of the data and thus the errors should be largely free of structure.  We show theoretically that mean and variance changes in these potentially sophisticated raw data structures will manifest in the forecast errors (under certain assumptions on the forecasting model). Furthermore we adapt Page's CUSUM detector (Section A) to detect the manifested changes in the forecast errors. The detection of a mean change in the forecast errors indicates that the forecasting model needs re-evaluating as it has become biased, while a variance change indicates that prediction intervals from the forecasting model will no longer be precise and would need to be re-constructed.

Let $\{Y_t:t=1,2,\ldots\}$ be a univariate time series \add{with $Y_t|y_{t-1},y_{t-2},\ldots$} following some unknown \add{conditional} distribution, $F$, with an associated forecasting model, $\mathcal{F}$. \add{We define a forecasting model as a principled approach to take in observed data $\{y_t\}$ and produce a prediction of the future data either in point form or as a distribution.}  Assuming a known forecasting model, $\mathcal{F}$, we define the point forecast for time point $t$ made $h$ time steps previously as,
\begin{equation}\label{eq:forecast}
	\hat{y}_{t}(h)=\EE{\mathcal{F}}{Y_{t}|y_{t-h},\ldots,y_1},
\end{equation}
\add{where $y_{t-h},\ldots,y_{1}$ are the previously observed realizations from the random process $Y_t$.}
Without loss of generality, we focus on one-step-ahead forecasts, hence $\hat{y}_t(1)$ is the forecast for time point $t$ made at time point $t-1$. Using \eqref{eq:forecast}, we define the forecast errors, $e_t$, as
\begin{equation*} \label{eq:forecastErrors}
	e_t=Y_t-\hat{y}_{t}(1).
\end{equation*}

First, we consider theoretically how mean changes in the potentially sophisticated raw data manifest in the forecast errors. We highlight the desirable properties of the forecast errors and describe how we can incorporate them into Page's CUSUM detector. Secondly, we consider the more general setting where mean and/or variance changes in the raw data can occur. Again, we show theoretically how these changes manifest in the forecast errors and describe an alternative adaption of Page's CUSUM detector to detect these changes.

\subsection{Mean Changes}
First, we consider how mean changes in potentially sophisticated raw data manifest in the forecast errors. Assume  there is a changepoint at some unknown location $k^*$,
\begin{align}
	Y_t|y_{t-1},\ldots,y_1&\sim F\;,\;\;\;\;\;\text{for }t=1,\ldots,m+k^*\;,\nonumber\\
	Y_t|y_{t-1},\ldots,y_1&\sim G\;,\;\;\;\;\;\text{for }t=m+k^*+1,m+k^*+2,\ldots\;,\label{eq:rawMeanModel}
\end{align}
where $F$ and $G$ differ in expectation,
\begin{equation}\label{eq:rawMeanChange}
	\EE{G}{Y_t|t>m+k^*,y_{t-1},\ldots,y_1}-\EE{F}{Y_t|t\leq m+k^*y_{t-1},\ldots,y_1}=\delta_{\mu,m}\neq0.
\end{equation}
To ease notation, we write $\tilde{y}_t$ to represent the whole past upon which $Y_t$ is dependent. Moreover, the relevant pre- and post-changepoint times are inferred from the distribution we are taking expectations with respect to. Note, \eqref{eq:rawMeanModel} encompasses the mean change model in (A1) from supplementary material
and additionally allows the raw data to have many different properties including, but not limited to, being temporally dependent, exhibiting trends, exogenous variables, or containing seasonality.

For the mean change in \eqref{eq:rawMeanChange} to manifest in the forecast errors, we must place the following assumptions on the forecasting model being used.

\begin{assumption}\label{ass:meanForecasting}
	We have a forecasting model, $\mathcal{F}$, such that
	\begin{equation*}
		\EE{F}{Y_t|\tilde{y}_t}-\EE{\mathcal{F}}{Y_t|\tilde{y}_t}=b_\mu\;\;\;\;\;\text{for }t=1,\ldots,m+k^*\;,
	\end{equation*}
	where $b_\mu$ is some unknown (potentially zero) bias in the forecasts.
\end{assumption}
\begin{assumption}\label{ass:meanForecasting2}
	For a forecasting model, $\mathcal{F}$, and data that follows \eqref{eq:rawMeanModel} then,
	\begin{equation*}
		\EE{G}{Y_t|\tilde{y}_t}-\EE{\mathcal{F}}{Y_t|\tilde{y}_t}=b_\mu+f(\delta_{\mu,m})=b_\mu+\Delta_{\mu,m}\;\;\;\;\text{for }t=m+k^*+1,m+k^*+2,\ldots,\;,
	\end{equation*}
	where $\Delta_{\mu,m}\neq0$ and is some function of the change size in the raw data.
\end{assumption}

For generality, these assumptions state that, in the pre-change regime, the forecasting model has a constant bias, $b_\mu$, which, in practice, would ideally be equal to zero. When a change in the raw data occurs, many forecasting models will adjust in some way to try and account for this change. For example, in an AR(1) model with autoregressive parameter, $\phi$, a change size of $\delta_{\mu,m}$ in the raw data would result in a change size of $\Delta_{\mu,m}=\delta_{\mu,m}(1-\phi)$ in the forecast errors after one time point; assuming a perfectly specified forecasting model. Assumption \ref{ass:meanForecasting2} states that, like the AR(1) example above, the forecasting model cannot fully adjust to the change in the raw data and some constant change size still exists. This seems intuitive because if the forecasting model fully adjusts to the changepoint then we would not want to signal a changepoint as the forecasting model has adapted to the change and is still performing well.

Under Assumptions \ref{ass:meanForecasting} and \ref{ass:meanForecasting2}, the change in mean in \eqref{eq:rawMeanModel} will result in a change in expectation in the forecast errors that is similar to the original Page-CUSUM (A1), that is,
\begin{equation*}\label{eq:meanModelErrors}
	Y_t-\hat{y}_t(1)=e_t=\begin{cases}
		b_\mu+\epsilon_{\mu,t},&t=1,\ldots,m+k^*,\\
		b_\mu+\epsilon_{\mu,t}+\Delta_{\mu,m},&t=m+k^*+1,m+k^*+2,\ldots\;,
	\end{cases}
\end{equation*}
with relevant null (\add{$H_0$:}$\Delta_{\mu,m}=0$) and alternative (\add{$H_A$:}$\Delta_{\mu,m}\neq0$) hypothesis formally defined in Supplementary (A2) and (A3) with $\Delta_m=\Delta_{\mu,m}$.

As the forecasting model accounts for the dependence in the raw data, the zero-mean random variables $\{\epsilon_{\mu,t}:t=1,2,\ldots\}$ will be i.i.d. \add{These $\epsilon_{\mu,t}$ are essentially the residuals after accounting for the bias in the forecast model.} Note, even if the forecasting model is misspecified, the general assumptions on $\{\epsilon_t:t=1,2,\ldots\}$ given in Supplementary Material Assumptions A and B are likely to hold.

To detect the manifested mean change in the forecast errors, we use Page's CUSUM detector defined in Supplementary Material (A4), on the forecast errors, that is,
\begin{align*}
	Q_\mu(m,k)&=\Sum{t=m+1}{m+k}e_t-\frac{k}{m}\Sum{t=1}{m}e_t\;,\nonumber\\
	D_\mu(m,k)&=\max\limits_{0\leq i\leq k}\left|Q_\mu(m,k)-Q_\mu(m,i)\right|\;,
\end{align*}
with corresponding stopping time,
\[
	\tau_{\mu,m}=\min\left\{k\geq1:D_\mu(m,k)\geq\hat{\sigma}_{\mu,m}c_{\mu,\alpha}g(m,k,\gamma)\right\}\;.
\]
\add{Here we distinguish from the original CUSUM detector, $D(m,k)$ defined on $Y_t$ and our CUSUM detector, $D_\mu(m,k)$ defined on the forecast errors.  The original $D(m,k)$ can be obtained by replacing \add{$\epsilon_t$} with $y_t$ in $D_\mu(m,k)$ as shown in the Supplementary Material.}

As the random variables, $\{\epsilon_t:t=1,2,\ldots\}$, are i.i.d, $\hat{\sigma}_{\mu,m}$, can be estimated using the traditional estimate of the standard deviation of the forecast errors within the Phase I (training) period. The weight function, $g(m,k,\gamma)$ is defined as in (A5) in the supplementary material and the critical constant, $c_{\mu,\alpha}$ can be derived from the limiting distribution of the stopping time \add{under both the null and alternative hypotheses, due to Corollaries 1 and 2 detailed in the supplementary material.  These corollaries demonstrate that as long as we have a forecasting model that satisfies the given assumptions, a mean change in potentially sophisticated raw data will manifest as mean changes in the forecast errors and will be detectable under our framework.
}

\subsection{Mean and Variance Changes}
Thus far we have considered detecting mean changes using Page's CUSUM detector. \cite{Inclan1994} extend this to detect variance changes by considering the CUSUM detector of the (centered) squared data. This can be viewed as searching for a mean change in the variance estimates of the data. Here we employ a similar idea and show that mean and/or variance changes in the raw data will manifest as mean changes in the (centered) squared forecast errors.

We extend the mean change model in \eqref{eq:rawMeanModel} to allow for mean and variance changes. 
\begin{definition}\label{eq:rawVarModel}
\add{Again, assume there is a changepoint at some unknown location, $k^*$,
\begin{align*}
	Y_t|y_{t-1},\ldots,y_1&\sim F\;,\;\;\;\;\;\text{for }t=1,\ldots,m+k^*\;,\nonumber\\
	Y_t|y_{t-1},\ldots,y_1&\sim G'\;,\;\;\;\;\;\text{for }t=m+k^*+1,m+k^*+2,\ldots\;,
\end{align*}
where $F$ and $G'$ differ in expectation and variance,
\begin{align*}
	\EE{G'}{Y_t|t>m+k^*,y_{t-1},\ldots,y_1}-\EE{F}{Y_t|t\leq m+k^*y_{t-1},\ldots,y_1}=\delta_{\mu,m}\;,\\
	\Var{G'}{Y_t|t>m+k^*,y_{t-1},\ldots,y_1}-\Var{F}{Y_t|t\leq m+k^*y_{t-1},\ldots,y_1}=\delta_{\xi,m}.
\end{align*}}
\end{definition}
Under Assumption \ref{ass:meanForecasting}, the change in mean and/or variance will result in a mean change in the squared forecast errors,
\begin{equation*}\label{eq:varModel2}
	\left(Y_t-\hat{y}_t(1)-b_\mu\right)^2=\left(e_t-b_\mu\right)^2=\begin{cases}
		b_\xi+\epsilon_{\xi,t},&t=1,\ldots,m+k^*,\\
		b_\xi+\epsilon_{\xi,t}+\Delta_{\xi,m},&t=m+k^*+1,m+k^*+2,\ldots\;.
	\end{cases}
\end{equation*}
where $b_\xi=\mathbb{E}[\epsilon_{\mu,t}^2]$ \add{is the bias that may remain in the forecasts} and $\{\epsilon_{\xi,t}:t=1,2,\ldots\}$ are zero-mean random variables. Again, due to the forecasting model accounting for the dependence structure in the raw data, $\{\epsilon_{\xi,t}:t=1,2,\ldots\}$, will be i.i.d.

The change size in the (centered) squared forecast errors, $\Delta_{\xi,m}$, can be shown to be a combination of the mean and/or variance change in the raw data.
\begin{proposition}{}\label{prop:varChangeSize}

	Let $\{Y_t:t=1,2,\ldots\}$ follow \add{the model in Definition} \ref{eq:rawVarModel} with $\delta_{\mu,m}=0$ and $\mathcal{F}$ be a forecasting model that satisfies Assumption \ref{ass:meanForecasting}. Then,
	\begin{equation}\label{eq:pureVar}
		\Delta_{\xi,m}=\Var{G'}{Y_t|\tilde{y}_t}-\Var{F}{Y_t|\tilde{y}_t}\;.
	\end{equation}
	Moreover, if $\delta_{\mu,m}\neq0$ and Assumption \ref{ass:meanForecasting2} holds, then
	\begin{equation}\label{eq:meanVar}
		\Delta_{\xi,m}=\Var{G'}{Y_t|\tilde{y}_t}-\Var{F}{Y_t|\tilde{y}_t}+\Delta_{\mu,m}^2\;.
	\end{equation}
\end{proposition}
\begin{proof}
	We start by noting that the change size, $\Delta_{\xi,m}$, is the difference between the expectation of the squared forecast errors assuming the post change distribution $G'$ and the pre-change distribution $F$. From this, we can show the equivalences given in Proposition \ref{prop:varChangeSize} by substituting relevant definitions of the forecast errors and using Assumption \ref{ass:meanForecasting}.
	\add{More formally, we can describe this as follows.  Note to ease notation we assume $Y_t$ is dependent on its past throughout and the relevant time-points (pre- or post-change) are clear from the distribution which the expectations are taken with respect to. We begin with the case where $\delta_{\mu,m}=0$,
	\begin{align*}
		\Delta_{\xi,m}=&\EE{G'}{(e_{t}-b_{\mu})^2}-\EE{F}{(e_t-b_{\mu})^2}\\
		=&\EE{G'}{\left(Y_{t}-\EE{\mathcal{F}}{Y_{t}}-b_{\mu}\right)^2}-\EE{F}{\left(Y_t-\EE{\mathcal{F}}{Y_t}-b_{\mu}\right)^2}\\
		=&\EE{G'}{\left(Y_{t}-\EE{F}{Y_{t}}\right)^2} -\EE{F}{\left(Y_{t}-\EE{F}{Y_{t}}+b_\mu-b_\mu\right)^2}\\
		=&\EE{G'}{\left(Y_{t}-\EE{G'}{Y_{t}}-\delta_{\mu,m}\right)^2} -\EE{F}{\left(Y_{t}-\EE{F}{Y_{t}}\right)^2}\\
		=&\Var{G}{Y_t}-\Var{F}{Y_t}
	\end{align*}
	If $\delta_{\mu,m}\neq0$ and Assumption \ref{ass:meanForecasting2} holds, then
		\begin{align*}
		\Delta_{\xi,m}=&\EE{G}{(e_{t}-b_\mu)^2}-\EE{F}{(e_t-b_\mu)^2}\\
		=&\EE{G}{\left(Y_{t}-\EE{\mathcal{F}}{Y_{t}}-b_\mu\right)^2}-\EE{F}{\left(Y_t-\EE{\mathcal{F}}{Y_t}-b_\mu\right)^2}\\
	=&\EE{G}{\left(Y_{t}-\EE{G}{Y_{t}}+b_\mu+\Delta_{\mu,m})-b_\mu\right)^2} -\EE{F}{\left(Y_{t}-\EE{F}{Y_{t}}+b_\mu-b_\mu\right)^2}\\
	=&\EE{G}{\left(Y_{t}-\EE{G}{Y_{t}}\right)^2+\Delta_{\mu,m}^2}-\EE{F}{\left(Y_{t}-\EE{F}{Y_{t}}\right)^2}\\
	=&\Var{G}{Y_{t}}-\Var{F}{Y_t}+\Delta_{\mu,m}^2\;.
	\end{align*}}
\end{proof}

To detect the manifested changes in the centered squared forecast errors, we can again adapt Page's CUSUM detector,
\begin{align*}
	Q_\xi(m,k)&=\Sum{t=m+1}{m+k}\left(e_t-\hat{b}_{\mu,m}\right)^2-\frac{k}{m}\Sum{t=1}{m}\left(e_t-\hat{b}_{\mu,m}\right)\;,\nonumber\\
	D_\xi(m,k)&=\max\limits_{0\leq i\leq k}\left|Q_\xi(m,k)-Q_\xi(m,i)\right|\;,
\end{align*}
with corresponding stopping time,
\[
	\tau_{\xi,m}=\min\left\{k\geq1:D_\xi(m,k)\geq\hat{\sigma}_{\xi,m}c_{\xi,\alpha}g(m,k,\gamma)\right\}\;.
\]
Here $\hat{b}_{\mu,m}$ is the mean estimate of the forecast errors in the $m$ Phase I (training) samples and $\hat{\sigma}_{\xi,m}$ is the traditional estimate of the standard deviation of the centered squared forecast errors within the Phase I (training) period. Again, $g(m,k,\gamma)$ is defined as in (A5) from the supplementary material and the critical constant $c_{\xi,\alpha}$ can be derived from the limiting distribution of the stopping time under $H_0$, which is known due to the following corollary.

\begin{corollary}{Following from Theorem 1 in Supplementary Material}\label{cor:varChangeH0}

	Let $\{Y_t:t=1,2,\ldots\}$ follow \add{the model in Definition} \ref{eq:rawVarModel} and $\mathcal{F}$ be a forecasting model that satisfies Assumption \ref{ass:meanForecasting}. Under $H_0$, the limit distribution in Theorem 1 from the supplementary material, holds for $D(m,k)=D_\xi(m,k)$ and $\hat{\sigma}_m=\hat{\sigma}_{\xi,m}$.
\end{corollary}
Using Proposition \ref{prop:varChangeSize}, we gain the limiting distribution of $\tau_{\xi,m}$ under $H_A$ from the following proposition.

\begin{proposition}{}\label{prop:varChangeHA}

	Let $\{Y_t:t=1,2,\ldots\}$ follow \add{the model in Definition} \ref{eq:rawVarModel} with $\delta_{\mu,m}=0$ and $\mathcal{F}$ be a forecasting model that satisfies Assumption \ref{ass:meanForecasting}. Furthermore, assume $\Delta_{\xi,m}$ in \eqref{eq:pureVar} from Proposition \ref{prop:varChangeSize} satisfies Assumptions C-E from the supplementary material. Then under $H_A$, the limit distribution in Theorem 2 from the supplementary material, holds with $\tau_m=\tau_{\xi,m}$ and $c_\alpha=c_{\xi,\alpha}$. Moreover, the limit distribution holds if $\delta_{\mu,m}\neq0$ and $\mathcal{F}$ also satisfies Assumption \ref{ass:meanForecasting2} with $\Delta_{\xi,m}$ in \eqref{eq:meanVar} from Proposition \ref{prop:varChangeSize} satisfying Assumptions C-E from the Supplementary Material.
\end{proposition}
\begin{proof}
	From Proposition \ref{prop:varChangeSize}, given a certain change size in the raw data, we know the change size in the forecast errors. This, in combination with Theorem 2 from the supplementary material, yields the above result.
\end{proof}

Propositions \ref{prop:varChangeSize} and \ref{prop:varChangeHA} and Corollary \ref{cor:varChangeH0} show that provided the mean and variance changes in the potentially sophisiticated raw data are large enough, they will result in a mean change in the squared forecast errors that is detectable using our framework.

\section{Common Forecasting Models}\label{sec:Examples}
Here we investigate some common forecasting models and show how these can be used within our framework to detect different types of changes. We consider ARMA and ETS forecasting models, showing when they satisfy our forecasting model assumptions and give two introductory simulation examples illustrating the effectiveness of their forecast errors for detecting changepoints. Note a more detailed simulation study is performed in Section \ref{sec:Simulation}.

\subsection{ARMA Models}
We denote $\{Y_t:t=1,2,\ldots\}$ by the ARMA$(p,q)$ process,
\begin{equation}\label{eq:ARMA}
	\phi(B)(Y_t-\lambda_t)=\theta(B)\epsilon_t,\;\;\;\;\;t=1,2,\ldots\;,
\end{equation}
where $\lambda_t$ are the mean parameters, $\phi(B)=1-\phi_{1}B-\ldots-\phi_{p}B^p$ and $\theta(B)=1+\theta_{1}B+\ldots+\theta_{q}B^q$ are the autoregressive and moving average polynomials respectively and $B$ is the back-shift operator. The error terms, $\epsilon_t$, are assumed to be independent zero-mean random variables with variance, $\sigma_t^2$. We assume the ARMA process is causal and invertible meaning,
\begin{equation*}
	\phi(x)\neq0\;\;\;\text{and}\;\;\;\theta(x)\neq0\;\;\;\;\;\text{for all }|x|\leq1\;.
\end{equation*}

Interest lies in testing whether the time-dependent parameters (specifically $\lambda_t$ and $\sigma_t$) remain constant or change at some time point. A change in $\lambda_t$ corresponds to a mean change in the data while a change in $\sigma_t$ corresponds to a variance change. Here we investigate how an ARMA forecasting model would react to changes in $\lambda_t$ and $\sigma_t$.

\begin{figure}[htbp]
	\centering
	\includegraphics[width=0.8\textwidth]{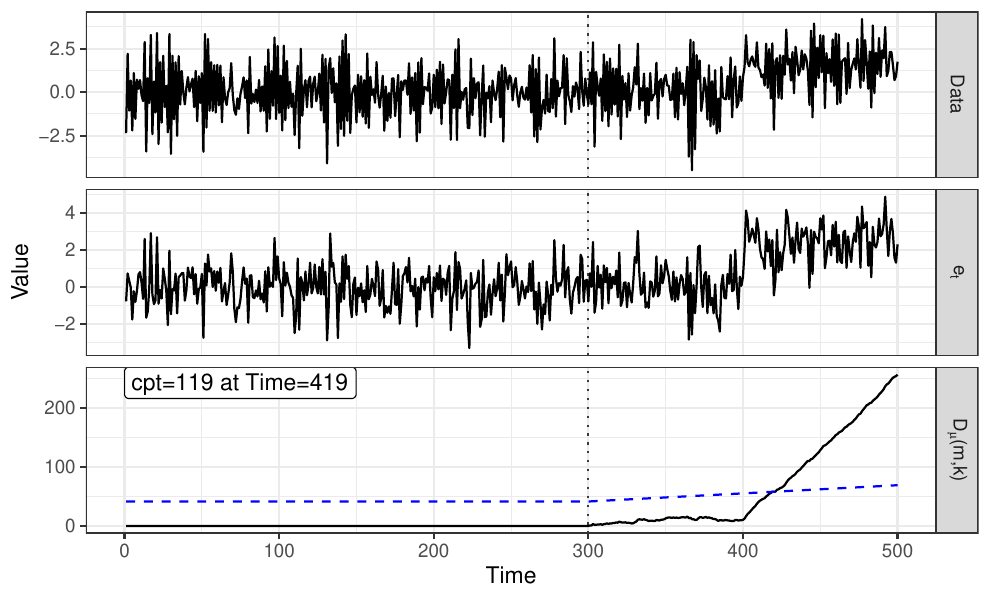}
	\caption{Data generated from an ARMA(1,1) model with a change in mean at time point 400; the resulting forecasting errors when using an ARMA forecasting model; and the detector, $D_{\mu}(m,k)$, where the dashed line shows the associated threshold for detecting a change. The vertical dotted line represents the start of the monitoring period.}
	\label{fig:ARMAexample}
\end{figure}

First, we present a small simulated example to display the effectiveness of using forecast errors to detect a mean change in an ARMA(1,1) model. For details on how the data was generated and the creation of the forecasting model see Section \ref{sec:Simulation}. Figure \ref{fig:ARMAexample} shows some raw data generated from an ARMA(1,1) model with $\lambda=0$, $\phi=-0.8$, $\theta=0.2$ and a change in mean at time point 400. Using an ARMA forecasting model and sequentially producing one-step-ahead forecasts we obtain the forecast errors shown. These forecast errors appear i.i.d and clearly display the changepoint in the raw data. Moreover, the detector (here $D_\mu(m,k)$ with $m=300$) rises sharply after the changepoint and hits the required threshold (the dashed line) to detect the change just 19 time points after it has occurred. Figure \ref{fig:ARMAexample} demonstrates that mean changes in ARMA models will manifest in the forecast errors. Additionally, as the forecast errors are i.i.d this makes detecting the changepoint easier using $D_\mu(m,k)$ than using $D(m,k)$ on the raw data. This is demonstrated further in Section \ref{sec:Simulation}.

Detecting changes in ARMA model parameters was also considered in \cite{Aue2015}. They proposed monitoring $\epsilon_t$ and $\epsilon_t^2$ as changes in the ARMA model parameters would manifest in $\epsilon_t$ and $\epsilon_t^2$. This approach is similar to ours, as under a perfectly estimated ARMA model the residuals and forecast errors would be identical. Note, the method in \cite{Aue2015} is exclusively for ARMA models where as our framework is more general and can be used with many different forecasting models.

In the mean change scenario, \cite{Aue2015} showed that a change size of $\delta_{\mu,m}$ in the mean parameter $\lambda$ would manifest in $\epsilon_t$ (and therefore our forecast errors) as
\begin{equation*}
	\Delta_{\mu,m}=\delta_{\mu,m}\left(1-\sum\limits_{j=1}^p\phi_j\right)\sum\limits_{l=0}^\infty\psi_l(\theta)\;,
\end{equation*}
where \[
	\frac{1}{\theta(z)}=\sum\limits_{l=0}^\infty\psi_l(\theta)\;.
\]
Therefore, Assumptions \ref{ass:meanForecasting} and \ref{ass:meanForecasting2} are satisfied and assuming $\Delta_{\mu,m}$ satisfies Assumptions C-E from the supplementary material, then Corollaries 1 and 2 hold, giving us the limit distribution of the stopping times under $H_0$ and $H_A$.

Similarly, for a change in variance in the data of size $\delta_{\xi,m}$, \cite{Aue2015} showed this would produce a change size in $\epsilon_t^2$ (and therefore our squared forecasting errors) as $\Delta_{\xi,m}=\delta_{\xi,m}$. Note, as expected this matches with Proposition \ref{prop:varChangeSize}.
Again, if $\Delta_{\xi,m}$ satisfies the Assumptions C-E from the supplementary material and $\Delta_{\mu,m}=0$ then Corollary \ref{cor:varChangeH0} and Proposition \ref{prop:varChangeHA} hold.

\subsection{ETS Models}\label{sec:ETSmodels}
ETS models cover a wide range of model structures. Here we show that generally ETS forecasting models do not satisfy Assumption \ref{ass:meanForecasting2} when detecting mean changes, however our methodology will still allow for the detection of variance changes. For demonstration purposes, for the mean change setting, we consider the local level model, ETS(A,N,N), with an additive error structure, no trend and no seasonality. We can produce one-step-ahead forecasts using the ETS(A,N,N) model by the recursive relation,
\begin{equation}\label{eq:ES}
	\hat{y}_t(1)=\hat{y}_{t-1}(1)+\alpha e_{t-1}\;,
\end{equation}
where $e_{t-1}$ is the forecast error at time point $t-1$ and $0<\alpha<1$ is the smoothing parameter. A larger value of $\alpha$ will cause the model to adapt to the new data more quickly, while a smaller value will place more weight on historical values making the model less sensitive to recent values.

ETS models adapt well to mean changes in raw data and in the majority of scenarios return to being unbiased after a few time steps as shown in the following proposition.
\begin{proposition}\label{prop:ETSmean}
	Let $\{Y_t:t=1,2,\ldots\}$ follow the change in mean model in \eqref{eq:rawMeanModel} and $\mathcal{F}$ be a forecasting model that follows \eqref{eq:ES}. Assuming $\mathbb{E}_F\left[Y_t|\tilde{y}_t\right]=\lambda$, the expectation of the forecast errors, after the changepoint, is
\[
	\Ee{e_{k^*+s}}=\begin{cases}
		\delta_{\mu,m}&\text{for }s=1\\
		(\lambda+\delta_{\mu,m})\left(1-\sum\limits_{i=0}^{s-2}\alpha(1-\alpha)^i\right)-\lambda\left(\alpha(1-\alpha)^{s-1}-(1-\alpha)^s\right)&\text{for }s=2,3,\ldots.
	\end{cases}
\]
\end{proposition}
\begin{proof}
	This follows from the repeated substitution of the previous forecast errors into \eqref{eq:ES} until the pre-change regime is reached.
\end{proof}
Proposition \ref{prop:ETSmean} shows that as $s$ increases the forecast errors will return to zero assuming $\alpha>0$. Hence, Assumption \ref{ass:meanForecasting2} is not satisfied unless $\alpha=0$. Moreover, we can see for larger $\alpha$ this convergence to zero will occur faster.

Thus mean changes in the raw data will only manifest in the forecast errors if $\alpha$ is small.  This is one of the advantages of the ETS model - it adapts to non-stationarity in the mean.  However, our framework can still be used to detect variance changes in the raw data. These variance changes will affect the prediction interval resulting from the ETS model. Here we show a small example illustrating how a variance change in raw data can make prediction interval poor and how this can be detected using the squared forecast errors from an ETS model.

\begin{figure}[htbp]
	\centering
	\includegraphics[width=0.8\textwidth]{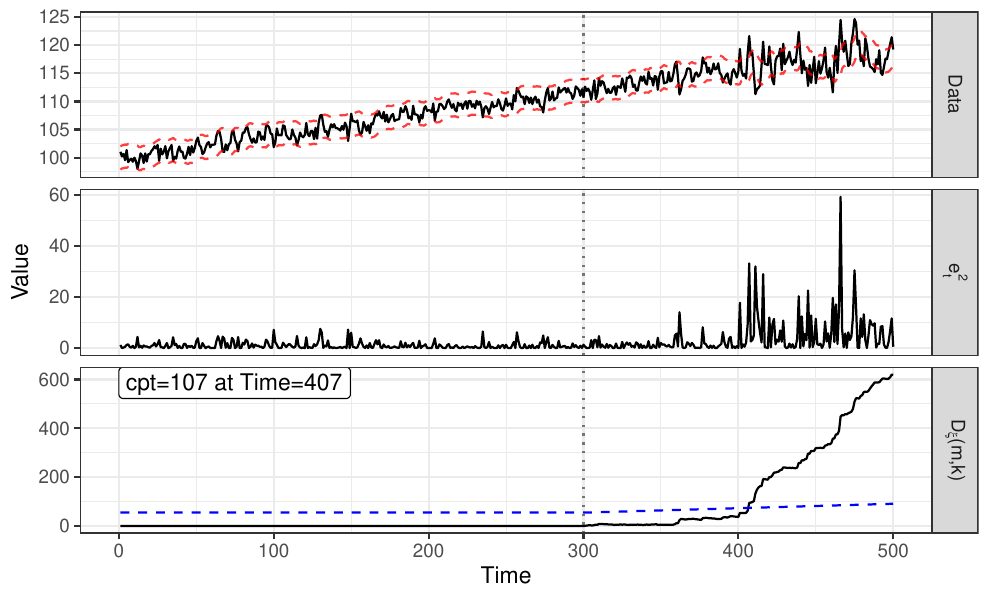}
	\caption{Data with a variance change at time point 400, with dashed lines showing prediction intervals from an ETS forecasting model. The squared forecast errors are shown along with the Detector with the dashed line being the threshold. The vertical dotted line shows the start of the monitoring period.}
	\label{fig:ETSexample}
\end{figure}

Figure \ref{fig:ETSexample} shows raw data that exhibits a trend and a variance change at time point $400$. We use an ETS(A,A,N) model to forecast this data and the prediction interval (based on $\pm2$ standard errors) are shown by the dashed lines. Clearly, after the changepoint there are many data points lying outside the prediction interval and hence we need to detect the change in variance so the prediction interval can be re-calibrated. The squared forecast errors show a clear mean change after the variance change in the data has occurred and the detector (here $D_\xi(m,k)$ with $m=300$) reaches the threshold, shown by the dashed line, just 7 time points after the change has occurred.

This mean change in the squared forecast errors is to be expected due to Proposition \ref{prop:varChangeSize}. Hence, assuming the variance change size in the raw data, $\delta_{\xi,m}$, satisfies Assumptions C-E from the supplementary material, then the limit distribution of the stopping time under $H_0$ and $H_A$ is given in Corollary \ref{cor:varChangeH0} and Proposition \ref{prop:varChangeHA}.

\section{Simulation}\label{sec:Simulation}
We now examine the performance of our framework in a selection of simulation examples. We present changes in mean, trend and autoregressive parameters within seasonal data.  Simpler model structures are given in the supplementary material.
Note that Assumptions \ref{ass:meanForecasting} and \ref{ass:meanForecasting2} required for the theoretical validity of the stopping rule are not necessarily satisfied in all scenarios, however, we still show satisfactory performance.

We compare the detector $D_\xi(m,k)$ with the squared forecast errors from an appropriate 1) ARMA and 2) ETS model against 3) the detector $D(m,k)$ with the (centered) squared raw data as in \cite{Inclan1994}. Throughout we will refer to $D(m,k)$ based upon the raw data as \textit{Raw CUSUM}; $D_\cdot(m,k)$ based upon the ARMA forecast errors as \textit{ARMA Forecast Errors}; and $D_\cdot(m,k)$ based upon the ETS model forecast errors as \textit{ETS Forecast Errors} with the choice of $D_\mu(m,k)$ or $D_\xi(m,k)$ dependent on the type of change.

The forecasting models are trained on an initial proportion of the data of length $n_\text{train}$\add{=300}. Then the forecasting models sequentially produce one-step-ahead forecasts yielding the required forecast errors. To generate the ARMA forecasting models we use the \textit{Arima} function within the \textit{forecast} R package \citep{Hyndman2021} and for the ETS forecasting model we use the \textit{es} function within the \textit{smooth} R package \citep{Svetunkov2021}. To determine when a change has occurred we use the theoretical thresholds from Theorem 1 from the supplementary material and Corollaries 1 and \ref{cor:varChangeH0} with $\gamma=0$ and false alarm rate set at 0.05. For the \textit{Raw CUSUM}, to estimate $\hat{\sigma}_m$ within in the threshold we use the Bartlett long run variance estimator \citep{Andrews1991} from the \textit{sandwich} R package \citep{Zeileis2020}. For the \textit{ARMA Forecast Errors} and \textit{ETS Forecast Errors}, we assume the forecast errors are i.i.d so the traditional standard deviation estimator is used. Throughout, we repeat each scenario 1000 times and report the average detection delay (ADD) with error bars showing 2 empirical standard errors either side of the mean; detection probability (DP) and false detection probability (FDP). The ADD is the mean of the detection delays given a change was signalled within the length of the data. The DP is the proportion of simulations where a change was detected within the length of the monitoring data. Finally, the FDP is the proportion of simulations where a change was detected before the true change had occurred. Hence, we seek an ADD as close to 0 as possible; a DP close to 1 and a FDP close to 0. Unless otherwise specified, we assume the random errors (innovations) are Normally distributed but note our framework does not require this assumption.

Throughout, we generate data with an ARMA error structure introduced in \eqref{eq:ARMA}. In practice, it is rare to know the exact data generating process and this raises questions regarding the misspecification of our forecasting model. Clearly, given the data generating process is ARMA, the most appropriate forecasting model should be an ARMA model with the same order.
However, we use the \textit{auto.arima} function from the \textit{forecast} package to pick the order of the ARMA model as well as estimating the model parameters. This means the forecasting model used to generate the \textit{ARMA Forecast Errors} may also be misspecified. We show that despite these misspecifications, a good forecasting model generally still produces better detection results than using the \textit{Raw CUSUM}.

\subsection{Mean and Trend Changes in Seasonal Data}
We examine two different scenarios, one where a mean shift occurs and one where a trend (drift) emerges in the underlying data that includes a seasonal component, to mimic applications. For the mean change scenario, we generate errors that follow an ARMA(2,1) model with $\phi=(-0.6, 0.3)$, $\theta=-0.3$, $\lambda=0$ and $\sigma^2=1$. We add a changepoint of size $\delta_\mu$ at time point $k^*=100$. We add a seasonal component to the data with frequency 12, mimicking monthly data. Within each season the data follows the curve $y=10\sin(x)$ with $x\in[0,\pi]$ so the mean of each seasonal component is taken from 12 equally spaced points along this curve.

For the trend change scenario, we simulate errors, $\epsilon_t$, following an ARMA(1,1) model with $\phi=0.2$, $\theta=0.2$, $\lambda=0$ and $\sigma^2=1$. We simulate the data with the form $y_t=\mathbbm{1}[t>m+k^*]\beta t+\epsilon_t$ for varying gradients $\beta$. Hence, the pre-change data has no trend and the post-change data has a trend with specified gradient, $\beta$. Moreover, we add a seasonal component to the data with frequency 7, mimicking daily data with day of the week effects. Each 7 time point cycle follows the curve $y=10\cos(x)$ with $x\in[0,2\pi]$, so the mean of each seasonal component is taken from 7 equally spaced point along this curve.

Due to the added seasonal component, we can no longer specify the exact models for the \textit{ARMA Forecast Errors}, without using dummy regressors or Fourier expansions. In practice, it is common to use the \textit{auto.arima} function from the \textit{forecast} package for the \textit{ARMA Forecast Errors}. We estimate the seasonal ARMA forecasting model, specifying the model should have no differencing. This function estimates the most appropriate seasonal ARMA model using AIC.

\begin{figure}[tbp]
	\centering
	\includegraphics[width=1\textwidth]{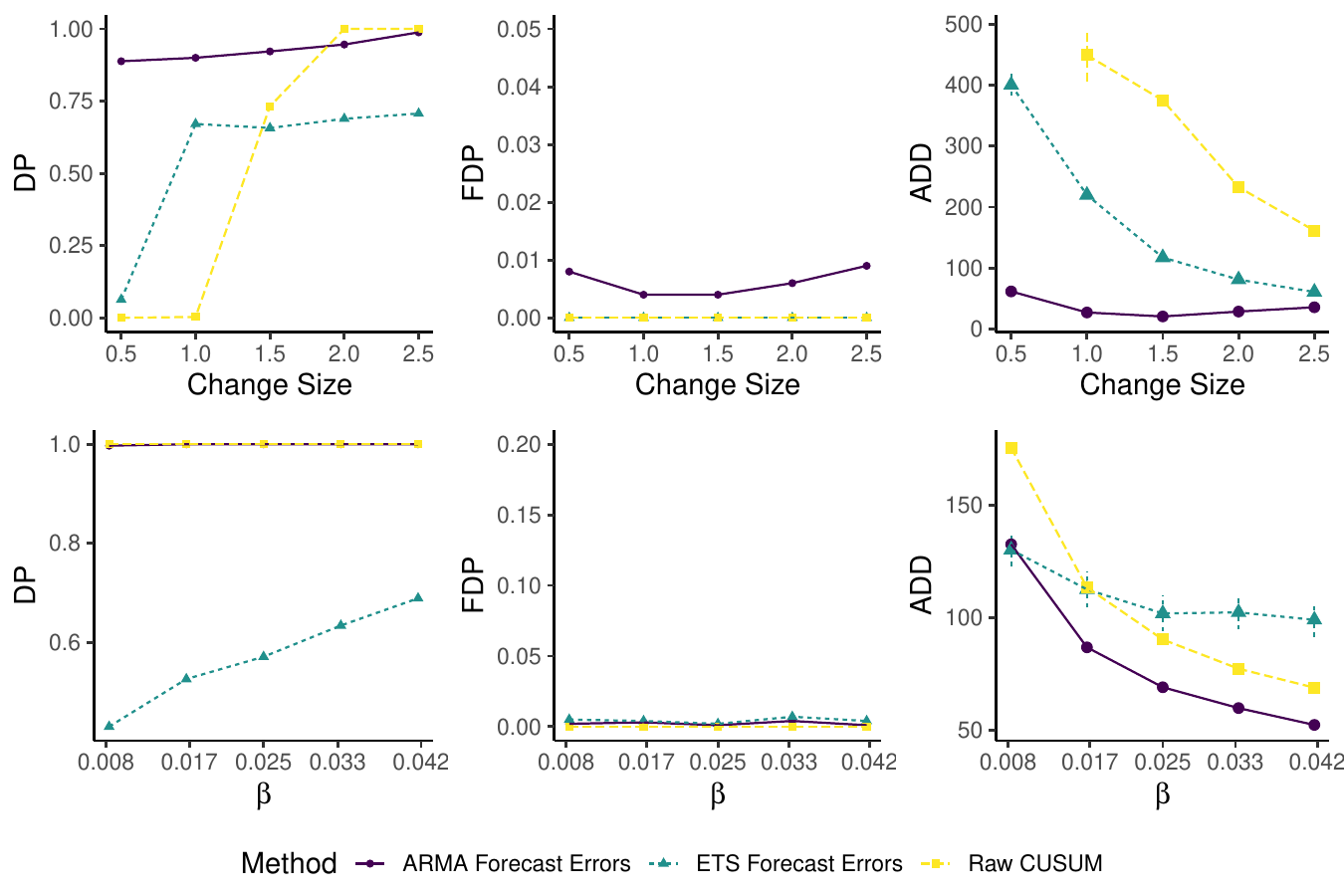}
	\caption{Detection Probability (DP), False Detection Probability (FDP) and Average Detection Delay (ADD), with error bars (barely visible) showing 2 standard errors either side of the mean, for two scenarios; top, a change in mean and bottom, a change in trend within seasonal data.}
	\label{fig:meanTrendSeason}
\end{figure}

Figure \ref{fig:meanTrendSeason} shows some interesting results. For the mean change scenario, \textit{ARMA Forecast Errors} has a high DP and extremely low ADD across all scenarios and outperforms the \textit{Raw CUSUM}. In the change in trend scenario, both \textit{ARMA Forecast Errors} and \textit{Raw CUSUM} have extremely high DP, however, the \textit{ARMA Forecast Errors} has a lower ADD across all change sizes. The FDP is low across all scenarios.

\subsection{Change in Autoregressive Parameter in Seasonal Data}\label{sec:arChange}
We now examine scenarios where we have data with AR(1) generated errors which undergoes a change in its autoregressive parameter at the specified changepoint. For this section, we include seasonality with a frequency of 4 mimicking quarterly data. The length of the data, training sizes etc remain the same as in the previous scenarios and we test two scenarios:
\begin{enumerate}
	\item The pre-change data has AR(1) errors with $\phi_{\text{pre}}=0.2$ and $\lambda=0$. The post-change errors remain AR(1) but with varying parameters, $\phi_{\text{post}}$.
	\item The pre-change data has AR(1) errors with $\phi_{\text{pre}}=0.8$ and $\lambda=0$. The post-change errors remain AR(1) but with varying parameters, $\phi_{\text{post}}$.
\end{enumerate}
The seasonality is created by adding dummy variables making the mean of each quarter $(-2,5,7,-10)$ respectively.  We use the \textit{auto.arima} function from the \textit{forecast} package to fit a seasonal ARMA model and we specify the model should have no differencing.  For the \textit{ETS Forecast Errors} we use an ETS(A,N,A) model with  smoothing parameters to be estimated.

\begin{figure}[tbp]
	\centering
	\includegraphics[width=\textwidth]{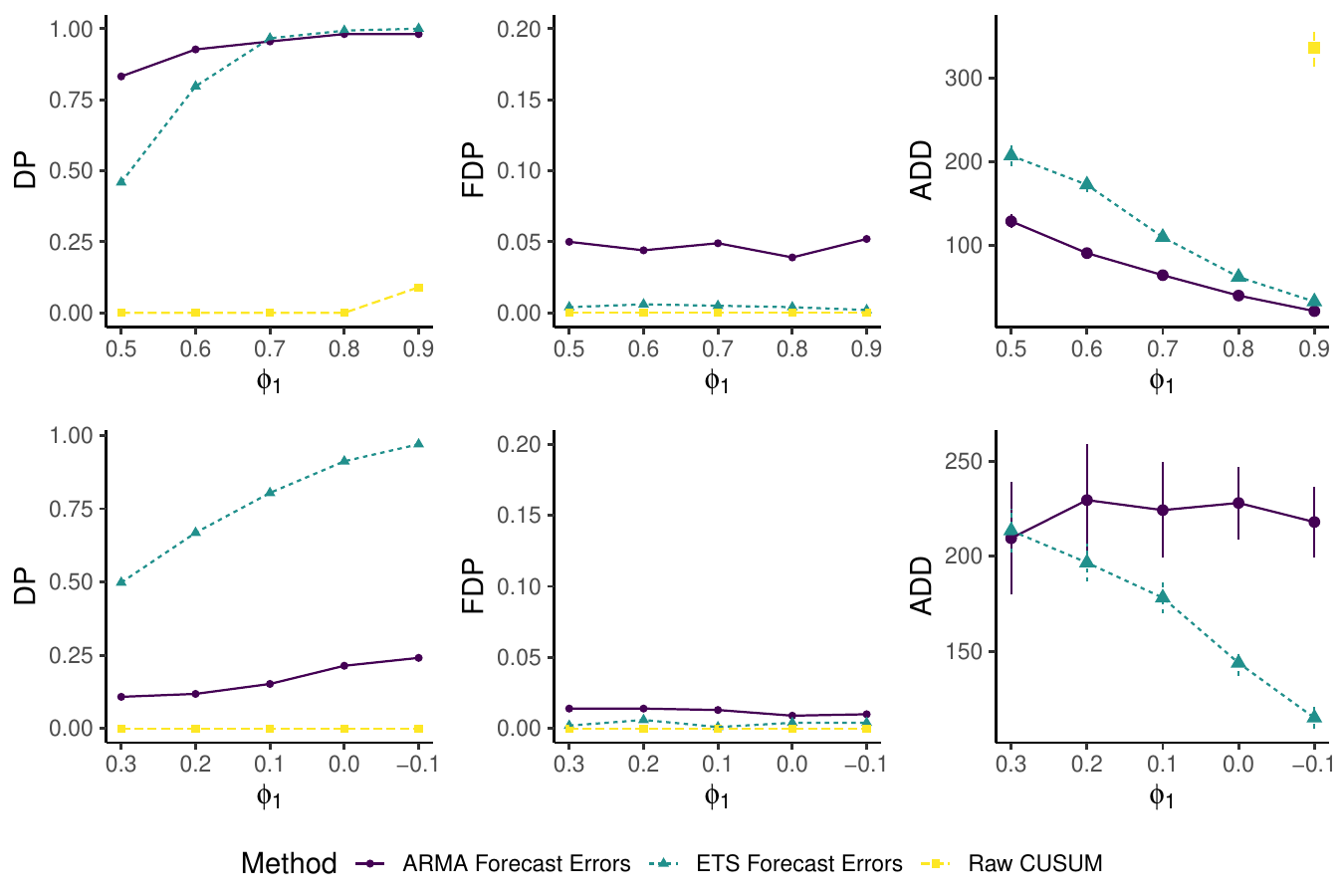}
	\caption{Detection Probability (DP), False Detection Probability (FDP) and Average Detection Delay (ADD), with error bars showing 2 standard errors either side of the mean, for two scenarios with a change in autoregressive parameter of different sizes; top $\phi_0$=0.2, bottom $\phi_0$=0.8.}
	\label{fig:ar}
\end{figure}

Figure \ref{fig:ar} shows that the DP and ADD is better for all methods when the dependence increases after the change. In all scenarios, using forecast errors gives better results than the \textit{Raw CUSUM}.  This is not unexpected as the \textit{Raw CUSUM} does not take the seasonality into account, which results in an overestimation of the variance in the Phase I (training) period.  A larger variance estimates makes changes much harder to detect. The FDP is conservative for all methods ($<0.075$).

\section{Application}\label{sec:Application}
We now apply this approach to two different applications. Firstly, we seek to determine a change in mail volumes indicating the start of the Christmas peak season within a Royal Mail delivery office by examining the forecast errors of the number of parcels being delivered each day. This is an example where only the forecast errors are known and the underlying data and forecasts are unknown to us. Secondly, we monitor NHS A\&E admissions for Gallstone related pathology (GRP) and seek to identify any changepoints by fitting an appropriate ARMA forecasting model.

\subsection{Parcel Delivery Volumes}
Each day Royal Mail forecast the number of parcels that need to be delivered from each delivery office across the UK. With accurate forecasts, the Delivery Office managers can make informed decisions for each day such as ensuring there are enough staff on a shift to meet demand. Each year the number of parcels being delivered increases around the Christmas period, however the exact date can differ from year to year and between different delivery offices. Hence, to maintain accurate staffing levels Royal Mail needs to quickly identify when the Christmas peak season begins.

Here we aim to identify the start of the Christmas peak season in one specific delivery office. We take the forecast errors received from Royal Mail, which date from the 1st July 2020 through to 31st December 2020. We perform our analysis assuming we receive the data in a sequential fashion and aim to identify the changepoint as soon as possible. The Christmas peak season can start anytime from the start of November, hence we will use the data from July to October as our Phase I (training) sample and begin monitoring the forecast errors from 1st November. We will use both detectors $D_\mu(m,k)$ and $D_\xi(m,k)$ with the theoretical thresholds from Corollaries 1 and \ref{cor:varChangeH0} to determine when a change has occurred.

\begin{figure}[htbp]
	\centering
	\includegraphics[width=0.8\textwidth]{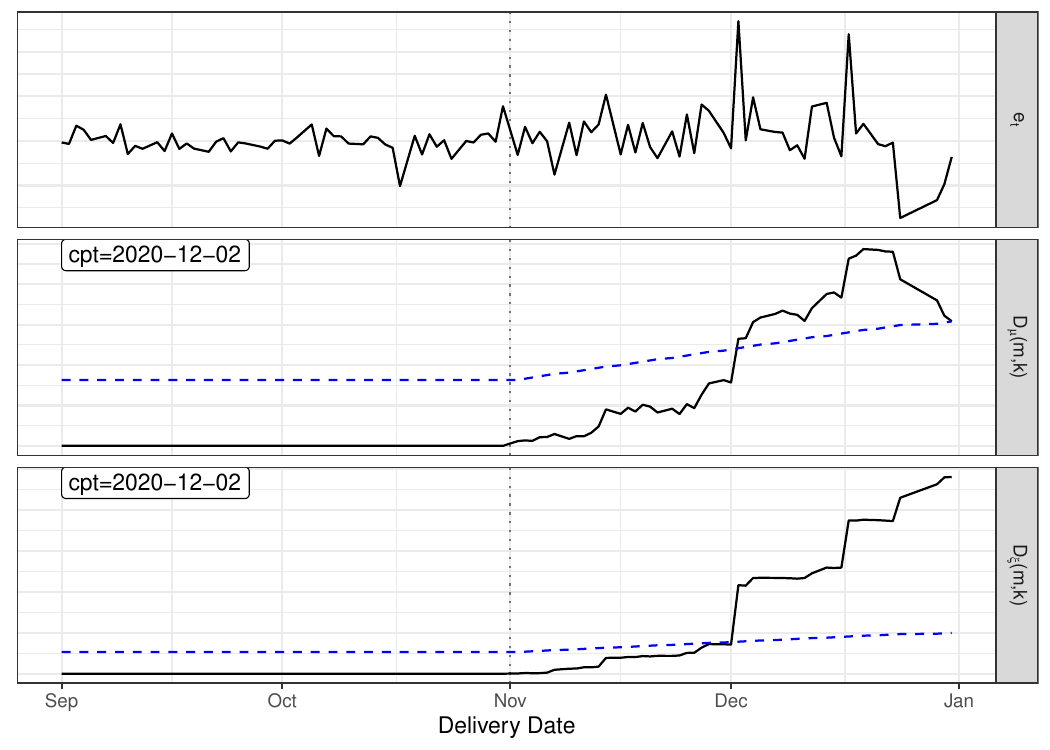}
	\caption{Forecast errors for the number of parcels to be delivered from a specific Royal Mail delivery office in 2020; the detctors $D_\mu(m,k)$ and $D_\xi(m,k)$ with associated thresholds shown by the dashed lines.}
	\label{fig:RoyalMail}
\end{figure}

Figure \ref{fig:RoyalMail} shows the forecast errors along with the detectors $D_\mu(m,k)$ and $D_\xi(m,k)$. Both the detectors identify a changepoint on the 2nd of December and looking at the forecast errors this corresponds to a rise in the mean and variance, indicating the start of the Christmas peak season.

Using this information Royal Mail would be able to adjust their forecasting model to account for the increase in parcels being delivered around the Christmas period in the knowledge that the Christmas peak season has begun.  These forecasts feed into weekly scheduling of staff thus avoiding staff shortages.

\section{GRP Admissions}
\begin{figure}[htbp]
	\centering
	\includegraphics[width=0.8\textwidth]{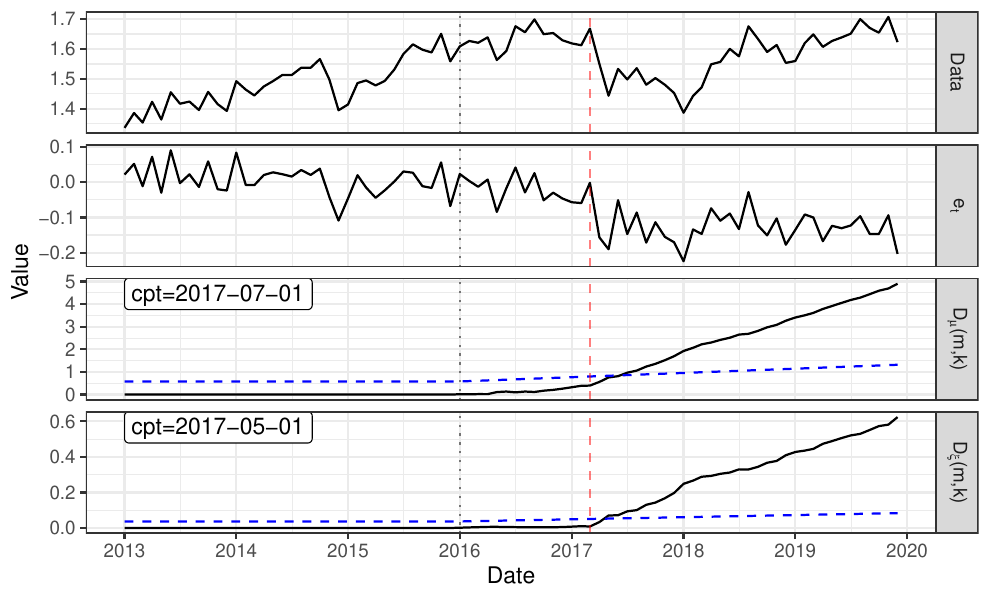}
	\caption{Data showing the proportion of GRP A\&E admissions; the forecast errors from a SARIMA$(1,0,0)(1,0,0)_{12}$ model; and the detectors $D_\mu(m,k)$ and $D_\xi(m,k)$ with associated thresholds shown by the horizontal dashed lines. The dotted line represents the start of the monitoring period and the vertical dashed line indicates a changepoint identified in the retrospective analysis.}
	\label{fig:gallstones2}
\end{figure}

We examine the proportion of A\&E admissions related to Gallstone related pathology (GRP) from a number of participating hospitals across England. This dataset is based on NHS Hospital Episode statistics and was initially analyzed in \cite{Taib2021}. The data consists of the monthly proportion of GRP A\&E admissions and is shown in the top panel of Figure \ref{fig:gallstones2}. There is a clear changepoint in the trend that we wish to identify that is marked by the vertical dashed line. The analysis of this data set in \cite{Taib2021} was performed retrospectively after all the data from 2010-2020 was available, however to allow for accurate forecasting, using a sequential changepoint method could detect this changepoint in a timely manner.

\cite{Taib2021} found that a SARMA$(1,0,0)(1,0,0)_{12}$ model was appropriate for the data with a time regressor to account for the trend. To identify a change in a sequential manner, using a model-based approach, would require at least two years worth of data to account for the yearly seasonality - this could be far too long and result in poor forecasts of GRP admissions.

We demonstrate the effectiveness of our method by fitting a SARMA$(1,0,0)(1,0,0)_{12}$ on the data up to 2013 and use this as our forecasting model. We used this forecasting model to generate one-step-ahead forecasts for the remaining data up to 2020. We used the data from 2013-2016 as our Phase I (training) data of size $m=36$ and performed our analysis using both detectors $D_\mu(m,k)$ and $D_\xi(m,k)$ with monitoring beginning in 2016.

Figure \ref{fig:gallstones2} shows the forecast errors and the detectors along with the associated theoretical thresholds from Corollaries 1 and \ref{cor:varChangeH0}. We can see that the changepoint is identified in $D_\mu(m,k)$ just four data points after a changepoint identified in \cite{Taib2021} (marked by the dashed line). Additionally, $D_\xi(m,k)$ identified the changepoint just 2 time points after the change occurred. Both of these detectors have a much shorter detection delay than using a model-based sequential changepoint algorithm which has a minimum possible detection delay of 24 months due to the yearly seasonality in the data.

\section{Conclusion}\label{sec:Conclusion}
This paper presents a new framework for monitoring forecasting models through the use of sequential changepoint detection. We have shown that common changepoints in underlying, potentially complex, data generating processes manifest in the one-step-ahead forecast errors under certain assumptions and can therefore be detected using sequential changepoint analysis. Moreover, we have shown by monitoring the forecast errors rather than the raw data for changepoints, we can greatly reduce the detection delay of changepoints while maintaining a low false positive rate. When using ARMA forecasting models, we have shown that changes in mean and variance in the underlying data can be detected. ETS models often adapt to mean changes, but can still be used to detect variance changes, which can cause prediction intervals to become miscalibrated. Applying our framework to delivery volume forecast errors from a Royal Mail delivery office allowed for the rapid detection of the Christmas peak season thus allowing future forecasts and staffing to be appropriately adjusted. Furthermore, we showed using a forecasting model on NHS A\&E admission data allowed us to detect a changepoint in the proportion of GRP admissions related in a shorter time frame than would be possible using retrospective analysis or a model-based sequential changepoint approach.

There are a number of extensions that could be made to this framework. Firstly, it is becoming increasingly common for multivariate forecasts to be made where many variables/products are forecasted at once. Incorporating a multivariate sequential changepoint algorithm on the resulting vector of forecast errors would be an interesting extension but raises additional questions such as the sparsity of the change and is therefore left as future work. Additionally, here we focused on one-step-ahead forecast errors but the framework is general, for $h$-steps ahead. Combining information from multiple forecast horizons would be an interesting extension.

\section*{Acknowledgements}
Grundy is grateful for the support of the Engineering and Physical Sciences Research Council (grant number EP/L015692/1) and Royal Mail Group Ltd. Killick is grateful for the support of the UK Research Councils through grants EP/T014105/1, EP/T021020/1, EP/R01860X/1 and NE/T012307/1.  We express thanks to Adnan Taib for insights into the data.

\bibliographystyle{apalike}
\bibliography{Bib}


\end{document}